\documentclass[10pt,twocolumn]{article}

\usepackage[T1]{fontenc}
\usepackage[utf8]{inputenc}
\usepackage[margin=1in,columnsep=0.25in]{geometry}
\usepackage{amsmath,amssymb}
\usepackage{graphicx}
\usepackage{booktabs}
\usepackage{tabularx}
\usepackage{array}
\usepackage{enumitem}
\usepackage{hyperref}
\usepackage{cite}
\usepackage{authblk}
\usepackage{titlesec}
\usepackage{abstract}
\usepackage{fancyhdr}
\usepackage{xcolor}
\usepackage{caption}
\usepackage{float}
\usepackage{multirow}
\usepackage{orcidlink}

\hypersetup{
  colorlinks=true,
  linkcolor=blue!60!black,
  citecolor=blue!60!black,
  urlcolor=blue!60!black
}

\titleformat{\section}{\normalfont\bfseries\large}{\thesection.}{0.5em}{}
\titleformat{\subsection}{\normalfont\bfseries}{\thesubsection.}{0.5em}{}
\titleformat{\subsubsection}{\normalfont\itshape}{\thesubsubsection.}{0.5em}{}
\renewenvironment{abstract}{%
  \small\quotation\noindent\textbf{Abstract---}\ignorespaces
}{\endquotation}

\begin{document}

\twocolumn[{%
\begin{center}
  {\LARGE\bfseries Operationalising Information Security Management:\\[4pt]
  A Procedural Framework Analysis of ISO/IEC 27001:2022\\[4pt]
  Implementation in a Financial-Technology Organisation}\\[12pt]

  {\large Ratul Ali\,\orcidlink{0000-0003-0460-6141}}\\[4pt]
  {\normalsize
   Senior Consultant (IT) $\cdot$
   IOTA $\cdot$ Dhaka, Bangladesh\\[2pt]
   \href{mailto:abdurrahimratulalikhan@gmail.com}{\texttt{abdurrahimratulalikhan@gmail.com}}
   \quad$\cdot$\quad
   \href{https://orcid.org/0000-0003-0460-6141}{\texttt{orcid.org/0000-0003-0460-6141}}
  }\\[10pt]
\end{center}

\begin{abstract}
Organisations operating within information-intensive environments face intensifying pressure to formalise the governance of information security. The ISO/IEC 27001:2022 standard provides a globally recognised framework for establishing, implementing, maintaining, and continually improving an Information Security Management System (ISMS). This article analyses the procedural architecture deployed in a financial-technology organisation's ISMS, examining eight core operational procedures: IT Risk Assessment and Treatment, User Code of Conduct, Password Policy, Access Control, Internet Access, Physical Security, Backup and Restore Management, and Nonconformity Root Cause Analysis and Corrective Action. Drawing on documented internal training materials, the article investigates how each procedure operationalises the requirements of Annex~A controls and Clauses~6--10 of ISO~27001:2022. The paper evaluates the CIA Triad as a unifying evaluation criterion, the twelve-step risk assessment methodology, role-based responsibility allocation, and the interplay between corrective action governance and continual improvement. The findings suggest that a tightly integrated, multi-layered procedural hierarchy, supported by clear accountability structures and measurable risk metrics, constitutes the foundation of an effective ISMS implementation in financial-technology operating environments.
\end{abstract}

\vspace{12pt}
}]

\section{Introduction}
\label{sec:intro}

Information security governance has evolved from a purely technical discipline into a board-level management responsibility. The proliferation of digital assets, cloud-hosted services, and remote-working arrangements has substantially expanded the attack surface of modern organisations, making structured security frameworks indispensable~\cite{iso27001_2022}. The ISO/IEC~27001:2022 standard, published by the International Organization for Standardization, remains the most widely adopted international benchmark for ISMS certification, providing a risk-based approach to selecting and implementing security controls across all organisational assets~\cite{nist_csf}.

This article examines the ISMS procedural framework of a financial-technology organisation that has implemented a system aligned with ISO~27001:2022 (hereafter referred to as ``the Organisation''). The Organisation's ISMS encompasses a hierarchical document structure comprising policies, procedures (Level-2 documents), work instructions, and supporting forms. This study focuses on the Level-2 procedural layer, which translates high-level policy intent into day-to-day operational guidance for all staff, contractors, and consultants who interact with organisational information assets.

The motivation for this analysis is twofold. First, empirical studies indicate that procedural ambiguity and role confusion are among the most frequently cited causes of ISMS nonconformities during third-party audits~\cite{disterer2013}. Second, the 2022 revision of ISO~27001 introduced restructured Annex~A controls—reduced from 114 to 93—making it timely to examine how practitioners operationalise these revised controls at the procedural level~\cite{iso27001_2022}. In the financial-technology sector, where data confidentiality and service availability are both commercially and regulatorily critical, such analysis yields insights with broad applicability.

The remainder of this article is structured as follows. Section~\ref{sec:background} reviews relevant literature on ISMS frameworks and risk assessment methodologies. Section~\ref{sec:framework} describes the Organisation's ISMS document hierarchy. Sections~\ref{sec:risk}--\ref{sec:nc} analyse each of the eight core procedures. Section~\ref{sec:discussion} synthesises cross-cutting themes, and Section~\ref{sec:conclusion} presents conclusions and future research directions.

\section{Background and Related Work}
\label{sec:background}

\subsection{ISMS Frameworks and Standards}

An Information Security Management System is defined under ISO~27001:2022 as the system by which an organisation manages risks related to the confidentiality, integrity, and availability of information~\cite{iso27001_2022}. The standard adopts the Plan-Do-Check-Act (PDCA) management cycle and aligns with the ISO High-Level Structure (HLS) to facilitate integration with complementary management systems such as ISO~9001 (quality) and ISO~22301 (business continuity)~\cite{beckers2015}.

The NIST Cybersecurity Framework (CSF) and COBIT~2019 offer complementary perspectives on information security governance~\cite{nist_csf,cobit2019}. While NIST CSF emphasises outcome-based functions—Identify, Protect, Detect, Respond, and Recover—ISO~27001 takes a requirements-based approach, mandating documented controls and explicit management commitment. The procedural framework examined in this study draws primarily on ISO~27001:2022 Annex~A but incorporates elements of risk quantification methodology consistent with NIST SP~800-30~\cite{nist_sp800}.

\subsection{Risk Assessment Methodologies}

IT risk assessment methodologies range from qualitative approaches, such as the OCTAVE Allegro model, to quantitative models like Factor Analysis of Information Risk (FAIR)~\cite{fair2020}. A common hybrid approach—employed extensively in financial-sector organisations—uses a five-point ordinal scale for both impact and likelihood, producing a numerical risk score that maps to predefined risk appetite thresholds~\cite{stoneburner2002}. The twelve-step risk assessment process examined in this article is consistent with this hybrid approach, explicitly evaluating assets against the CIA Triad and calculating residual risk after control application.

\subsection{Organisational Behaviour and ISMS Compliance}

Research by Bulgurcu et al.~\cite{bulgurcu2010} demonstrates that user intention to comply with ISMS policies is strongly influenced by cost-benefit perceptions and normative beliefs. This underscores the importance of clear, role-differentiated procedures—such as a well-designed User Code of Conduct—alongside technical controls. Disterer~\cite{disterer2013} further argues that ISO~27001 certification alone is insufficient without a mature corrective action culture, highlighting the criticality of the nonconformity management process examined in Section~\ref{sec:nc}. These findings motivate the structured analysis presented in this article.

\section{ISMS Document Hierarchy and Procedural Scope}
\label{sec:framework}

The Organisation's ISMS is structured as a four-tier document hierarchy: Level-0 (ISMS Manual and Scope Statement), Level-1 (Policies), Level-2 (Operational Procedures), and Level-3 (Work Instructions and Supporting Forms). This architecture is consistent with best-practice guidance for ISO~27001 documentation and facilitates audit traceability~\cite{humphreys2008}. Table~\ref{tab:procedures} catalogues the eight Level-2 procedures that form the analytical focus of this study.

\begin{table}[H]
\centering
\caption{Level-2 ISMS Operational Procedures Analysed}
\label{tab:procedures}
\small
\renewcommand{\arraystretch}{1.25}
\begin{tabularx}{\columnwidth}{@{}cX@{}}
\toprule
\textbf{No.} & \textbf{Procedure Title} \\
\midrule
P-01 & IT Risk Assessment \& Risk Treatment \\
P-02 & User Code of Conduct \\
P-03 & Password Policy \\
P-04 & Access Control Procedure \\
P-05 & Internet Access Procedure \\
P-06 & Physical Security Procedure \\
P-07 & Backup \& Restore Management \\
P-08 & Nonconformity RCA \& Corrective Action \\
\bottomrule
\end{tabularx}
\end{table}

The Information Security Department holds overarching responsibility for ISMS governance, while line departments—IT Operations, IT Risk Management Desk, and Internal Audit—execute specific procedural obligations. The Chief Information Security Officer (CISO) acts as the central coordination authority, ensuring inter-procedural consistency and managing the corrective action programme.

\section{IT Risk Assessment and Risk Treatment (P-01)}
\label{sec:risk}

\subsection{Objectives and Scope}

The IT Risk Assessment and Treatment procedure aims to: (i) strengthen the organisational risk management framework in a systemic manner; (ii) establish a formal and repeatable risk assessment process; (iii) cultivate a risk-based security culture across all departments; and (iv) introduce structured techniques for identifying, measuring, mitigating, and monitoring IT risks under formal policies. The assessment is carried out annually and on demand whenever significant changes occur to the IT environment.

\subsection{The CIA Triad as an Evaluation Framework}

All asset evaluations are anchored to the CIA Triad:

\begin{itemize}[leftmargin=1.2em,itemsep=2pt]
  \item \textbf{Confidentiality} — restricts access to information; unauthorised disclosure may result in legal liability, loss of public confidence, or organisational embarrassment.
  \item \textbf{Integrity} — ensures that information remains trustworthy and accurate; uncorrected unauthorised modification can lead to fraud, erroneous decisions, or regulatory violations.
  \item \textbf{Availability} — guarantees reliable access to information by authorised personnel; denial of access degrades mission effectiveness and causes loss of productive time.
\end{itemize}

Assets are classified into six groups—People, Procedure, Data/Information, Software, Hardware, and Networking—and each is evaluated against the three CIA dimensions at Low, Medium, or High levels of impact.

\subsection{The Twelve-Step Risk Assessment Process}

The twelve-step methodology provides a structured workflow from team formation through to formal risk owner approval.

\textbf{Steps 1–4 (Scoping and Asset Valuation).} A cross-functional Risk Assessment Team is established, comprising the Information Security Official (Team Leader), Head of IT, an IT Operations representative, and an IT Audit official. The assessment boundary is formally defined to encompass all IT system components—servers, routers, switches, firewalls, administrators, documentation, and business records—and submitted to senior management for approval. Assets are then identified, grouped by category, assigned to owners and custodians, and given a quantified criticality value on a five-point scale (1~=~Low to 5~=~Critical). Critical assets include core transactional servers, network switches, routers, and firewalls; dependencies between assets are also identified at this stage.

\textbf{Steps 5–7 (Threat and Vulnerability Identification).} Threats are categorised as natural (floods, earthquakes, storms), human (unintentional acts and deliberate attacks such as hacking, malicious software, and unauthorised access), or environmental (prolonged power failure, chemical or liquid exposure). Vulnerabilities are identified through reviews of security procedures and Vulnerability Assessment reports. All existing and planned safeguards are catalogued to inform subsequent impact and likelihood analysis.

\textbf{Steps 8–10 (Risk Calculation).} Impact is computed as a function of asset value and estimated business loss, rated on the same five-point scale. Likelihood is assessed based on threat frequency and the effectiveness of existing controls. The composite risk score is calculated as $R = f(\text{Impact},\,\text{Likelihood})$, producing a value in the range 1–100. Table~\ref{tab:riskmatrix} presents the risk rating scale and associated management response requirements.

\begin{table}[H]
\centering
\caption{Risk Rating Scale and Management Response}
\label{tab:riskmatrix}
\small
\renewcommand{\arraystretch}{1.25}
\begin{tabularx}{\columnwidth}{@{}llXl@{}}
\toprule
\textbf{Score} & \textbf{Rating} & \textbf{Decision Basis} & \textbf{Timeline} \\
\midrule
1–19   & Negligible  & Risk Appetite      & 12 months \\
20–39  & Minor       & Risk Tolerance     & 6 months  \\
40–59  & Moderate    & Mgmt.\ Notify      & 3 months  \\
60–79  & Significant & Mgmt.\ Trigger     & 1 month   \\
80–100 & Severe      & Immediate Action   & 1 month   \\
\bottomrule
\end{tabularx}
\end{table}

An overall portfolio risk health indicator is also applied: below 8\% is considered Strong; 8–16\% Satisfactory; 16–25\% Fair; 25–30\% Marginal; and above 30\% Unsatisfactory.

\textbf{Steps 11–12 (Residual Risk and Approval).} Residual risk—the risk remaining after applying security controls and process improvements—is calculated separately from the base risk, because not all risks can be fully eliminated in a dynamic technology environment. The residual risk is formally communicated to the Risk Owner (electronically or in hard copy), and approval is obtained before the assessment cycle is closed.

\subsection{Risk Treatment Strategies}

Four canonical treatment responses are available. \textbf{Accept}: used for low-risk assets where no immediate safeguards are required and the risk falls within the organisation's appetite. \textbf{Reduce}: implement operational, procedural, physical, personnel, and technical controls to ensure no critical downtime; employed when risks are assessed as high. \textbf{Transfer}: shift the risk burden externally via insurance, contracts, or outsourcing. \textbf{Avoid}: change or discontinue the activity or asset giving rise to the risk, applied when no other viable option exists. Treatment timelines range from one month (Urgent/Severe) to twelve months (Low priority, subject to management discretion).

\subsection{Risk Monitoring, Reporting, and the Risk Register}

Risk profiles are monitored continuously. Specific monitoring activities include periodic review of operational risk profiles and material loss exposures, assessment of the quality and appropriateness of mitigating actions, and documentation of each monitoring step for effective review. Formal reports are submitted to the Management Review Meeting and must be comprehensive, accurate, consistent, and actionable. They address critical and emerging risks, risk events and intended remedial actions, control effectiveness, and areas of imminent risk crystallisation. A centralised Risk Register consolidates all risk information, enabling management oversight and coordinated cross-departmental risk response.

\section{User-Facing Security Procedures (P-02, P-03, P-04)}
\label{sec:user}

\subsection{User Code of Conduct (P-02)}

The User Code of Conduct defines obligations and prohibitions applicable to all employees, consultants, and contractors interacting with the Organisation's IT infrastructure. Six compliance pillars are articulated: (i) professional and ethical conduct—acting with honesty, trust, fairness, and integrity at all times; (ii) confidentiality and diligence in handling all information assets; (iii) risk awareness and personal accountability for adverse security impacts; (iv) prevention of conflicts of interest; (v) adherence to all regulatory and legal requirements; and (vi) upholding service standards that reflect professional values.

Workstation security requirements mandate screen-lock activation when unattended, prohibition on credential sharing, restriction on unauthorised configuration changes, regular password rotation, and mandatory antivirus scanning of the local hard disk. Physical document management requires that printed materials containing sensitive information be stored under lock and key and shredded after use; employees must never leave written or printed information unattended in public areas or on desktops.

Device management obligations require that all allocated devices remain under personal physical control at all times. Theft or loss must be reported immediately to the IT Service Desk without delay. All organisation-owned mobile devices (laptops, tablets, mobile phones) must be password-protected to prevent unauthorised data access. Devices may not be removed beyond organisational premises without CISO authorisation, and all devices must be returned upon termination of employment.

The Acceptable Use Policy distinguishes permitted activities—business-purpose system use, reporting identified security weaknesses, using authorised portable storage after pre-use scanning—from prohibited activities, which include sharing login credentials, engaging in harassment, downloading or installing unauthorised software, and forwarding proprietary information without authorisation.

Internet usage is restricted to official business purposes. Specific content restrictions prohibit retrieval or transmission of discriminatory, obscene, defamatory, or threatening material and chain letters. Software downloads require IT Service Desk approval and CISO authorisation. Unauthorised devices must not be attached to production, staging, or development networks; guest devices are segregated to a dedicated guest network; and production network security keys must never be disclosed to unauthorised parties.

Social engineering awareness is embedded within the Code of Conduct, emphasising the risks of pretexting, phishing, and unsolicited credential requests. The Clear Desk and Clear Screen policy mandates that workstations be locked when unattended and that sensitive documents never be left visible on desks, in printers, or in other communal spaces.

\subsection{Password Policy (P-03)}

The Password Policy operationalises ISO~27001:2022 Annex~A control A.5.17 (Authentication information). Passwords serve as the primary authentication mechanism and are subject to rigorous construction and lifecycle requirements. Users are required to change passwords regularly, maintain strict confidentiality over credentials, and refrain from sharing passwords under any circumstances—including with IT staff. Antivirus scanning is coupled with password governance to ensure that credential-harvesting malware is detected before authentication data can be exfiltrated. Administrative and privileged accounts are subject to enhanced password requirements commensurate with the sensitivity of the systems they protect.

\subsection{Access Control Procedure (P-04)}

Access control governs which identities may access which information resources under which conditions, directly addressing ISO~27001:2022 Annex~A controls A.5.15–A.5.18. The procedure establishes role-based access provisioning: access rights are determined by documented business need and approved by the relevant department head in conjunction with the Information Security Department. Periodic access reviews detect and revoke stale or excessive entitlements. Privileged access to administrative systems is subject to a dedicated approval workflow and heightened oversight. The procedure also governs account lifecycle management—provisioning, modification, and timely de-provisioning upon role change or departure—ensuring that the principle of least privilege is maintained throughout.

\section{Infrastructure Security Procedures (P-05, P-06)}
\label{sec:infra}

\subsection{Internet Access Procedure (P-05)}

The Internet Access Procedure establishes governance over outbound and inbound internet traffic for all users and network segments. Internet access is classified as a business tool and its use is subject to continuous monitoring. Specific prohibitions include accessing non-business websites during working hours, downloading executable files without authorisation, and connecting unauthorised devices to production or staging networks. These controls correspond to ISO~27001:2022 Annex~A controls A.8.20 (Network security) and A.8.23 (Web filtering). The procedure further requires that the security keys of internal networks be obtainable only through the IT Service Desk, preventing unauthorised lateral movement by guest or contractor devices.

\subsection{Physical Security Procedure (P-06)}

Physical security constitutes a critical control layer that complements logical access governance. The procedure covers data centre and server room access control, visitor management, equipment protection, and environmental controls. Physical access permissions are aligned with logical access entitlements to ensure that individuals cannot gain physical proximity to systems they are not authorised to access logically. This defence-in-depth posture aligns with ISO~27001:2022 Annex~A controls A.7.1–A.7.14 (Physical and environmental security). Environmental controls address temperature, humidity, power continuity, and fire suppression, while equipment maintenance schedules are documented and tracked to prevent hardware failures that could compromise availability.

\section{Backup and Restore Management (P-07)}
\label{sec:backup}

\subsection{Introduction and Objectives}

The Backup and Restore Management Procedure protects all electronic information assets from accidental deletion, corruption, or system failure, and ensures timely restoration of systems and business processes. Its scope encompasses all databases, systems, and network devices within the organisational data centre, under the stewardship of the Information Security Department. Core objectives include: regular automated backups to secure, verified storage media; disaster recovery and business continuation capability; defined minimum retention requirements; documented media disposal procedures; and accommodation of special backup requirements identified through technical risk analysis.

\subsection{System and Network Device Backup Procedures}

\textbf{System backups} comprise a full Known Good State snapshot of all servers—including mail, application, and mission-critical systems—taken before any configuration change and stored on tape. All backup copies are automatically transferred to the Disaster Recovery (DR) site. System-level script backups are produced on any system-level change, and a stakeholder-approved rollback plan is maintained at all times. The Information Security Department maintains a separate backup of cryptographic keys. Employees are responsible for their personal email backup; no official data may be stored in personal cloud storage.

\textbf{Network device backups} capture complete device configurations after every successful implementation, and running and startup configurations are immediately backed up following any routing, VPN, or other configuration change. Backups are transferred to DR immediately after each change, and a quarterly baseline snapshot is maintained for all network devices. Restore or rollback is initiated upon device failure or misconfiguration.

\subsection{Backup Retention Matrix}

The retention policy is tiered by asset category, backup frequency, and business criticality, as shown in Table~\ref{tab:retention}. After the expiry of any retention period, media is returned to the Information Security Department for secure reuse or destruction.

\begin{table}[H]
\centering
\caption{Illustrative Backup Retention Periods}
\label{tab:retention}
\small
\renewcommand{\arraystretch}{1.25}
\begin{tabularx}{\columnwidth}{@{}llX@{}}
\toprule
\textbf{Asset Category} & \textbf{Frequency} & \textbf{Retention Period} \\
\midrule
Core Databases     & Daily / Monthly / Yearly & 1 Week / 1 Year / 10 Years \\
Mail Server (DB)   & Daily                    & 6 Months \\
All Other Servers  & Annually                 & 1 Year \\
Network Devices    & On-demand / Half-Yearly  & 7 Days / 6 Months \\
\bottomrule
\end{tabularx}
\end{table}

\subsection{Recovery Objectives and Backup Verification}

A Recovery Point Objective (RPO) of 48 hours and a Recovery Time Objective (RTO) of 48 hours are formally defined, subject to DR site availability, data volume, and infrastructure capacity. Backup verification entails regular log review for errors and anomalous job durations, monthly random test restores by each responsible team, and formal documentation of all test restore outcomes. Records of log reviews and test restores are retained for audit compliance. A Restoration Report—signed by the DBA and Head of IT—is produced for every restore operation and retained for future audit.

Restoration requests are submitted via the IT Service Desk and require Head of IT approval. Restoration is performed on demand and at a minimum frequency of once every six months to validate recovery capability. The process is fully documented for each individual server and application, and both on-site and off-site restoration processes must be captured in separate procedural documents.

\subsection{Media Management and Secure Disposal}

Backup media is classified into three tiers: Red (Confidential), Yellow (Internal Use Only), and Green (Public). All removable media must be encrypted where possible, scanned for malware upon connection to any external system, and transported under management authorisation via a secure, trackable courier. Media must never be left unattended at any stage during transport, and all outbound media movements are formally logged.

Disposal requires written confirmation from the responsible technical team, followed by physical destruction (drilling, crushing, or electromagnetic degaussing) that renders data irretrievable. A Disposal Certificate attesting to secure destruction must be issued upon completion. All obsolete, disposed, or reused items are recorded in a dedicated disposal register, satisfying audit trail requirements consistent with ISO~27001:2022 Annex~A control A.8.10 (Deletion of information).

\section{Nonconformity Root Cause Analysis and Corrective Action (P-08)}
\label{sec:nc}

\subsection{Introduction and Scope}

The Nonconformity Root Cause Analysis (RCA) and Corrective Action procedure provides the feedback mechanism through which deviations from ISMS requirements are systematically investigated, remediated, and permanently closed. It directly implements the requirements of Clause~10.2 (Nonconformity and corrective action) of ISO~27001:2022 and applies to all organisational work products, processes, services, employees, and contractors.

\subsection{Definitions}

Three foundational concepts underpin the procedure:

\begin{itemize}[leftmargin=1.2em,itemsep=2pt]
  \item \textbf{Nonconformity (NC)} — any nonfulfillment of a documented ISMS requirement.
  \item \textbf{Root Cause Analysis (RCA)} — a set of analytical and problem-solving techniques targeted at identifying the actual underlying cause of a nonconformity.
  \item \textbf{Corrective Action (CA)} — the action taken to permanently resolve the identified issue and prevent its recurrence.
\end{itemize}

\subsection{Sources and Reporting of Nonconformities}

Any organisational personnel may identify and must report nonconformities. Sources include complaint investigations (internal and external), process failures, internal ISMS audits, management reviews, and any other originating source. Reporting is standardised through a Nonconformity Report Form submitted electronically to the Corrective Actions Administrator. Security events that do not constitute formal nonconformities are handled under the separate Incident Management Procedure.

\subsection{The Seven-Step Corrective Action Process}

When a nonconformity requires a change to processes, controls, or documentation, the Change Management Process is formally invoked. The seven documented steps are:

\begin{enumerate}[leftmargin=1.5em,itemsep=2pt]
  \item Clearly describe the nonconformity or issue requiring corrective action.
  \item Take immediate steps to contain the problem and mitigate its consequences.
  \item Identify root causes in collaboration with the relevant department, addressing deficiencies in system design or implementation.
  \item Department management plans, designs, and implements corrective actions, including engineering modifications if required.
  \item Review proposed corrective action recommendations via the Risk Assessment Process and document the review outcome.
  \item Identify and implement the appropriate corrective actions to address, remedy, and eliminate the root cause.
  \item CISO verification of closeout and confirmation of corrective action effectiveness.
\end{enumerate}

\subsection{Key Governing Principles and Timeframes}

Three overarching principles govern all corrective actions. \textbf{Risk Assessment First}: proposed actions must undergo a risk assessment prior to implementation, particularly where they introduce new or modified controls that could themselves generate unforeseen risks. \textbf{Immediate Action}: consequences of incidents must be addressed without delay, with priority placed on restoring customer confidence and service integrity. \textbf{Proportionality}: the scope and depth of corrective action must be commensurate with the severity of the nonconformity and its assessed ISMS risk exposure.

All corrective actions carry a target completion deadline of \textbf{90 days}. If this deadline cannot be met, the CISO and management must be notified by email with a documented justification for the extension. Procedure exclusions require a Dispensation Form approved by Top Management; any nonfulfillment without prior dispensation is escalated for disciplinary action under the HR Procedure for ISMS. Action plans may be used to schedule and track open corrective actions. Once fully implemented, the Corrective Action Requirement form is signed off by the CISO following due verification, closing the PDCA loop.

\section{Discussion}
\label{sec:discussion}

\subsection{Integration and Coherence of the Procedural Framework}

The eight procedures examined collectively operationalise a comprehensive set of ISO~27001:2022 Annex~A controls spanning the organisational, people, physical, and technological domains. A notable strength of the framework is explicit role assignment at each procedural level, ensuring unambiguous accountability. The CISO serves as a cross-cutting authority with direct involvement in password governance, corrective action closure, backup exception approvals, and access control escalations. This centralised coordination model reduces the risk of procedural gaps at the interfaces between departments.

The CIA Triad provides a consistent evaluation lens across all eight procedures. Risk assessment scores are anchored to CIA impact values; access control decisions are driven by confidentiality and integrity requirements; backup policies are fundamentally motivated by availability objectives; and the corrective action process is initiated precisely when any CIA dimension is found to be inadequately controlled. This coherence strengthens the overall security posture by ensuring that all procedural elements reinforce the same foundational objectives.

\subsection{Twelve-Step Risk Assessment: Strengths and Limitations}

The twelve-step risk assessment methodology represents a mature, hybrid quantitative-qualitative approach. Explicit asset valuation on a five-point scale, combined with structured threat categorisation and vulnerability sourcing from formal VA reports, reduces subjective bias and improves reproducibility across annual assessment cycles. The Risk Register serves as a living document, enabling longitudinal trend analysis and evidence of continual improvement for third-party auditors.

A potential limitation is the inherently subjective nature of the likelihood scale, which relies on expert judgement regarding threat frequency. Without empirical threat intelligence data feeds, assessors may exhibit anchoring bias, particularly for novel threat vectors such as supply-chain attacks or zero-day exploits. Future iterations of the methodology could benefit from integration with real-time threat intelligence platforms to ground likelihood estimates in observed and verifiable data~\cite{nist_sp800}.

\subsection{Backup and Recovery Governance}

The 48-hour RPO and RTO targets reflect a risk-appetite decision that balances the operational cost of more frequent backup cycles against the acceptable magnitude of data loss. In sectors with more stringent regulatory requirements—such as Basel III for banking operations—tighter RPO targets of four hours or less may be mandated, and organisations should evaluate whether current targets remain adequate as they scale. The tiered retention matrix, retaining critical database backups for up to ten years, demonstrates alignment with long-term audit and regulatory obligations. The segregation of media classification into three tiers (Red, Yellow, Green) introduces a data-centric security perspective consistent with ISO~27001:2022 Annex~A control A.5.12 (Classification of information), ensuring that disposal procedures are proportionate to data sensitivity.

\subsection{Corrective Action Culture and Continual Improvement}

The 90-day target for corrective action closure, combined with CISO-verified effectiveness reviews, establishes a structured feedback loop that drives continual ISMS improvement. The requirement for a risk assessment prior to implementing any corrective action is particularly noteworthy: it prevents the inadvertent introduction of new vulnerabilities during the remediation process—a risk that is frequently overlooked in less mature ISMS implementations~\cite{disterer2013}. The requirement to formally modify ISMS procedures and training programmes wherever corrective actions reveal systemic gaps ensures that lessons learned are institutionalised rather than episodic.

\subsection{Human Factors and Compliance Behaviour}

The User Code of Conduct, Password Policy, and social engineering awareness components collectively acknowledge that technical controls alone are insufficient to achieve ISMS objectives in a people-dependent operating environment. Consistent with the findings of Bulgurcu et al.~\cite{bulgurcu2010}, this framework embeds compliance obligations in clearly communicated, role-specific procedures supported by proportionate disciplinary consequences for nonfulfillment. The explicit articulation of prohibited behaviours alongside permitted ones reduces ambiguity and strengthens normative compliance pressure. This behavioural governance layer is especially critical in financial-technology environments where insider threat, credential phishing, and social engineering remain the dominant attack vectors.

\section{Conclusion}
\label{sec:conclusion}

This article has presented a comprehensive analysis of the procedural framework underpinning an ISO~27001:2022 ISMS implementation in a financial-technology operating environment. The study demonstrates that effective ISMS operationalisation requires far more than policy documentation: it demands a coherent hierarchy of interoperable procedures, each grounded in the CIA Triad, each assigning unambiguous roles and responsibilities, and each connected to a feedback mechanism that drives continual improvement.

The twelve-step risk assessment process provides a replicable, quantifiable methodology for systematic threat and vulnerability management. The user-facing procedures establish the human-factors governance layer without which technical controls are routinely circumvented by insider negligence or deliberate misuse. The backup and restore framework ensures business continuity resilience with formally defined RPO and RTO targets, auditable media management, and verifiable test restore practices. The corrective action procedure closes the PDCA loop, ensuring that identified weaknesses translate into lasting institutional improvements rather than cyclically recurring audit findings.

Future research should examine the empirical effectiveness of these procedures through longitudinal audit data, the integration of automated threat intelligence into the risk assessment cycle, and the organisational change management practices that sustain ISMS maturity over time. Additionally, comparative studies across multiple ISO~27001-certified financial-technology organisations would yield generalisable insights into best-practice procedural design principles applicable to the wider sector.

\section*{Acknowledgments}

The author gratefully acknowledges the ISMS training and procedural documentation materials that served as the primary empirical basis for this study. Gratitude is also extended to the IOTA for the guidance and support throughout this research.


\end{document}